\documentclass[amssymb,twocolumn,aps,nofootinbib]{revtex4}
\usepackage{amsmath}
\usepackage{enumerate}
\begin{document}
\title{Decomposition of the Total Momentum in a Linear
Dielectric into Field and Matter Components}
\author{Michael E. Crenshaw}
\email{michael.e.crenshaw4.civ@mail.mil} 
\thanks{NOTICE: this is the author's
version of a work that was accepted for publication in Annals of Physics.
Changes resulting from the publishing process, such as peer review, editing,
corrections, structural formatting, and other quality control mechanisms may
not be reflected in this document. Changes may have been made to this work
since it was submitted for publication. A definitive version was subsequently
published in Annals of Physics, {\bf 338,} 97 (2013).}
\affiliation{US Army Aviation and Missile Research, Development, and Engineering Center, Redstone Arsenal, AL 35898, USA}
\date{\today}
\begin{abstract}
The long-standing resolution of the Abraham--Minkowski electromagnetic
momentum controversy is predicated on a decomposition of the total
momentum of a closed continuum electrodynamic system into separate field
and matter components.
Using a microscopic model of a simple linear dielectric, we derive
Lagrangian equations of motion for the electric dipoles and show that
the dielectric can be treated as a collection of stationary simple
harmonic oscillators that are driven by the electric field and produce a
polarization field in response.
The macroscopic energy and momentum are defined in terms of the
electric, magnetic, and polarization fields that travel through the
dielectric together as a pulse of electromagnetic radiation.
We conclude that both the macroscopic total energy and the macroscopic
total momentum are entirely electromagnetic in nature for a simple
linear dielectric in the absence of significant reflections.
\end{abstract}
\maketitle
%\ocis{260.2110,260.2065}
%\pacs{42.70.-a,78.20.Ci}
\par
\section{Introduction}
\par
In a dielectric environment, the momentum of the electromagnetic field
becomes a contested issue.
The matter was first raised over a century ago when
Minkowski \cite{BIMin}, in 1908, and Abraham \cite{BIAbr}, in 1909,
produced conflicting expressions for the energy--momentum tensor of an
electromagnetic field in a linear optical medium.
While a number of experiments seemed to validate the
Abraham momentum density
\begin{equation}
{\bf g}_A= \frac{1}{c}\left ( {\bf E}\times{\bf H}\right )
\label{EQvs.1.01}
\end{equation}
other experiments suggested that the Minkowski momentum density
\begin{equation}
{\bf g}_M=\frac{1}{c} \left ( {\bf D}\times{\bf B}\right ) 
\label{EQvs.1.02}
\end{equation}
provides the correct description of the momentum of light in a linear
medium \cite{BIBrev}.
The inability to decide between the two energy--momentum tensors and
their underlying momentum densities became known as the
Abraham--Minkowski controversy.
The resolution of the Abraham--Minkowski dilemma was provided by
Penfield and Haus \cite{BIPenHau}, based on earlier
work by M\o ller \cite{BIMol}, who showed that the issue is
undecidable because neither the Abraham momentum nor the Minkowski
momentum is the total momentum.
Instead, each corresponds to an open description of the system in which
the momentum is allocated differently between the field and the
matter \cite{BIBrev,BIPenHau,BIMol,BIKran,BISal,BIPfei}.
\par
Following the work of Penfield and Haus \cite{BIPenHau},
Gordon \cite{BIGord}, in 1973, assumed the Abraham form for the field
momentum subsystem and derived the material contribution to the momentum
in a dilute vapor by integrating the Lorentz force on each dipole.
Gordon derived the total momentum density 
\begin{equation}
{\bf g}_G= \frac{1}{c}\left ( n {\bf E}\times{\bf B}\right ) 
\label{EQvs.1.03}
\end{equation}
for a vapor that is sufficiently dilute that reflections can be
neglected.
Three years later, Peierls \cite{BIPei} followed a similar 
procedure to obtain a total momentum density
\begin{equation}
{\bf g}_P= \frac{n^2+1}{2c}\left ( {\bf E}\times{\bf B}\right )
\label{EQvs.1.04}
\end{equation}
that is the arithmetic mean of the Abraham and Minkowski forms.
The Gordon and Peierls momentum densities are equivalent to 
lowest order in $\delta n=n-1$.
The literature on the subject is extensive and the reader is referred to
reviews of the prior work by
Pfeifer, {\it et al.} \cite{BIPfei},
Barnett and Loudon \cite{BIcankin2},
Milonni and Boyd \cite{BIMilB},
and Kemp \cite{BIKemp} and the resource letter by
Griffiths \cite{BIGriffRL}.
\par
At this point in time, there is a strong consensus in the scientific
literature that only the total energy and the total momentum have
physical meaning.
Then any decomposition of the total momentum into a field
component and a matter component is valid as long as their sum is the
total momentum \cite{BIPfei}, but the separate momentum
components are not physical.
Nevertheless, current practice is to acknowledge that consensus and
then proceed to assume some form of field momentum, usually Abraham,
Minkowski, or both, and then add some momentum for the matter, usually
in the form of the movement of massive particles of the medium that are
forced into motion by being either pulled or pushed by the Lorentz
force.
The recent experiments of She, Yu, and Feng \cite{BISYF}, for example,
are analyzed in terms of an inward push force as light exits a nanometer
silica filament.
In contrast, earlier experiments by Ashkin and Dziedzic \cite{BIAshD}
indicate an outward pull force on the surface as light enters a liquid.
Pfeifer, Nieminen, Heckenberg, and Rubinsztein-Dunlop \cite{BIPfei}, and
also Ramos, Rubilar, and Obukhov \cite{BIRRObuk}, start with the
Abraham tensor for the electromagnetic field and add a dust tensor for
the material and claim that the result is the total energy--momentum
tensor.
By definition, the total energy--momentum tensor can
also be derived by starting with the Minkowski tensor and adding a
different material tensor.
Barnett \cite{BIBarn} identifies the Abraham momentum as the kinematic
momentum of the block of dielectric, as a whole, and the Minkowski
momentum as the canonical momentum that is associated with the movement
of massive particles embedded in the dielectric.
Barnett goes on to say that the total momentum is both the Abraham
momentum with a corresponding partial momentum and the Minkowski
momentum with a different partial momentum, which is again true by
definition of total momentum.
Mansuripur \cite{BIMani} treats the kinematic momentum of partially
reflecting matter, pushed by a radiation pressure associated with the
Fresnel reflection, together with the forward traveling light momentum.
Kemp \cite{BIKemp1} formalizes a subsystem approach for the field and
matter components of momentum.
\par
The most rigorous approach to light propagation in a material is to use
a microscopic model in which the microscopic electromagnetic field
interacts with individual electric dipoles in the vacuum.
If the distance between dipoles in the material is much smaller than the
wavelength, then the theory can be considerably simplified by applying
an averaging scheme to a volume containing many atoms per cubic
radiation wavelength.
Gordon \cite{BIGord}, for example, spatially averages the Lorentz force
on a single electric dipole to create a macroscopic Lorentz force
density.
Integrating the Lorentz force density, Gordon obtains a
{\it definite} formula
\begin{equation}
{\bf g}_{mat}= \frac{n-1}{c}\left ( {\bf E}\times{\bf B}\right ) 
\label{EQvs.1.05}
\end{equation}
for the material component of momentum density in the dielectric.
Gordon assumes the Abraham form for the momentum density of the field,
then the field plus material momentum is the total momentum
\begin{equation}
{\bf G}_{total}=
\int_{\sigma}\frac{n}{c}\left ({\bf E}\times{\bf B}\right ),
\label{EQvs.1.06}
\end{equation}
where integration of the momentum density is performed over a
volume $V$ that includes all
fields present that is extended to all space $\sigma$.
%Gordon's microscopic derivation of the material component of momentum
%is both incorrect and incomplete.
Saldanha \cite{BISal} and
Bradshaw, Shi, Boyd, and Milonni and Boyd \cite{BIBrad},
among others, have revisited Gordon's derivation of the total momentum
as the sum of the Abraham momentum for the field and a material momentum
arising from the Lorentz force on dipoles and derived the Minkowski
momentum, which is not conserved, for the total momentum.
In recent work \cite{BICB}, global conservation principles were applied
to a closed system consisting of a quasimonochromatic pulse of
electromagnetic radiation normally incident on a stationary simple
dielectric with refractive index $n$ through a gradient-index
antireflection coating.
Conservation of the total momentum in a thermodynamically closed system
with complete equations of motion \cite{BIKran} is decisive.
Then, despite problems with the original derivation,
Eq.~(\ref{EQvs.1.06}) is the proven total momentum for the light pulse
plus stationary dielectric closed system \cite{BICB,BISPIE}.
\par
In this article, we investigate the interaction of the electromagnetic
field with a simple linear dielectric that is modeled microscopically as
a collection of individual electric dipoles composed of a positively
charged particle and a negatively charged particle.
The work of Penfield and Haus \cite{BIPenHau} and of
Gordon \cite{BIGord} is reproduced in part, but we go on to show that
the dielectric can be treated in the usual and familiar way as a 
collection of stationary simple harmonic oscillators driven by the 
electric field.
Because the contribution to the total momentum occurs through the
resulting polarization field, the total momentum can be regarded as
electromagnetic in character.
We begin, in Section II, with an analysis of the macroscopic picture
of the total momentum as being composed of arbitrary field and matter
components of momentum.
At the level of continuum electrodynamics, we take the material
momentum to be the difference of the total momentum, which is
conserved, and a field momentum that is to be determined.
The temporal derivative of the material momentum density is the
macroscopic force density that can be expressed in terms of the fields
and a macroscopic material property.
Assuming that a decomposition of the total momentum into field and
material components exists, we find that each choice of the field
momentum results in a specific relation between the refractive index
and the macroscopic material property.
Then, the decomposition of the total momentum into field and material 
components is not arbitrary or degenerate, but unique and definite.
In Section III, we summarize Gordon's derivation of the macroscopic
Lorentz force from the microscopic Lorentz force.
We discuss the problem that occur when we use the macroscopic Lorentz
force, along with similar results by Peierls \cite{BIPei},
Saldanha \cite{BISal}, and
Bradshaw, Shi, Boyd, and Milonni \cite{BIBrad}, to obtain the
field momentum.
Finally, in Section IV, we adopt a microscopic model of the
material as being composed of discrete dipoles formed from individual
pairs of positive and negative charges.
Writing the microscopic Lagrangian in a center-of-mass coordinate system,
we derive Lagrange equations of motion using both the center-of-mass
coordinates and the relative coordinates.
The microscopic Lorentz force used by Gordon \cite{BIGord},
Penfield and Haus \cite{BIPenHau}, and other
authors can be extracted from the Lagrange equations for the
center-of-mass coordinates.
However, macroscopic movement of massive particles is not allowed
in a solid block of dielectric.
Setting the velocity of the center-of-mass of the dipoles to zero in the
Lagrange equations for the relative coordinates, we find the 
elementary and expected result that the dielectric can be treated as a
collection of stationary simple harmonic oscillators driven by the
electric field \cite{BIMVK}.
The polarization field that is generated by the oscillators 
contributes to both the total energy and the total momentum through
the refractive index $n$.
We conclude that, in the absence of reflections, the total
energy and the total momentum are electromagnetic in nature and do
not separate into field and material motion components.
\par
\section{Field and matter components of the total momentum}
\par
A physical theory is an abstract mathematical model of some limited
aspect of the real world \cite{BIRindler}, in this case, light
traveling at speed $c/n$ in a simple linear dielectric.
Our model dielectric consists of a rectangular prism of space in which
light travels at speed $c/n$, where $n$ is a real constant.
No such idealization of a dielectric exists in the real world.
Instead, the usual conditions for results of the model to correspond to 
results in the real world apply:
a simple dielectric is linear, isotropic, homogeneous, transparent,
and dispersion-negligible in a regime in which electrostrictive effects
and magnetostrictive effects can be neglected \cite{BIMikura}.
We assume that a quasimonochromatic optical pulse impinges on
the dielectric from the vacuum at normal incidence in the
plane-wave limit and that the dielectric is covered with a thin
gradient-index coating in order to make reflections and surface
radiation pressure negligible.
\par
The Abraham momentum is a common, but not exclusive, choice for the
momentum of the field in a
dielectric \cite{BIGord,BIMilB,BIMani,BIPfei}.
Rather than make this an {\it ansatz}, we start with a more general form
of the field momentum in a dielectric
\begin{equation}
{\bf G}_{field}= \int_{\sigma}
\frac{\zeta(n)}{c}\left ( {\bf E}\times{\bf B}\right ) dv,
\label{EQvs.2.01}
\end{equation}
with $\zeta(n)$ an unspecified function of $n$,
that covers most of the proposed forms of electromagnetic
momentum, ({\it e.g.}, the Abraham, Minkowski, and Peirels momentums).
We note that $\zeta$ is specifically {\it not} a function of the
electric and magnetic fields.
In addition, $\zeta$ is not arbitrary, but a definite function of 
macroscopic property constants that is to be determined.
The momentum of the material ${\bf G}_{mat}$ is the difference of the 
total momentum, Eq.~(\ref{EQvs.1.06}) and field
momentum, Eq.~(\ref{EQvs.2.01}).
Then
\begin{equation}
{\bf G}_{mat}={\bf G}_{total}-{\bf G}_{field}=
\int_{\sigma} \frac{n-\zeta(n)}{c}({\bf E}\times{\bf B})dv.
\label{EQvs.2.02}
\end{equation}
The material momentum density
\begin{equation}
{\bf g}_{mat}= \frac{n-\zeta(n)}{c}({\bf E}\times{\bf B})
\label{EQvs.2.03}
\end{equation}
leads directly to a macroscopic force density
\begin{equation}
{\bf f}_{mat}=\frac{\partial{\bf g}_{mat}}{\partial t}=
(n-\zeta(n)) \left ( 
\frac{\partial {\bf E}}{\partial (ct)} \times{\bf B}+
{\bf E}\times\frac{\partial {\bf B}}{\partial (ct)}
\right ).
\label{EQvs.2.04}
\end{equation}
We define a macroscopic material property
\begin{equation}
\xi_e(n)=n-\zeta(n)
\label{EQvs.2.05}
\end{equation}
that relates the electric and magnetic fields to a macroscopic force
density
\begin{equation}
{\bf f}_{mat}=
\xi_e \left ( 
\frac{\partial {\bf E}}{\partial (ct)} \times{\bf B}+
{\bf E}\times\frac{\partial {\bf B}}{\partial (ct)}
\right )
\label{EQvs.2.06}
\end{equation}
on the spatially distributed particles of matter.
Then the decomposition of the total momentum into field and matter
components,
\begin{equation}
{\bf G}_{field}=
\int_{\sigma} \frac{n-\xi_e}{c}({\bf E}\times{\bf B})dv
\label{EQvs.2.07}
\end{equation}
\begin{equation}
{\bf G}_{mat}=
\int_{\sigma} \frac{\xi_e}{c}({\bf E}\times{\bf B})dv,
\label{EQvs.2.08}
\end{equation}
is uniquely defined by macroscopic material properties $n$ and $\xi_e$.
\par
Since the 1960s, the resolution of the Abraham--Minkowski controversy
has been that the total momentum can be arbitrarily separated into field
and material components.
%This lack of definiteness has led to some rather loose arguments for
%the causation of contrary experimental results.
What has been proved here is that, {\it if} the total momentum can be
separated into a component for the field and a component for the motion
of matter driven by a Lorentz force, then there is a specific form
for each of these components in terms of macroscopic material property
constants.
\par
\section{Microscopic Treatment of Material Momentum}
\par
The total momentum of a closed system with complete equations of motion
is unique by virtue of its conservation.
It has been argued that the decomposition of the total momentum into
field and material is  arbitrary
\cite{BIBrev,BIPenHau,BIMol,BIPfei,BIGord,BIPei,BIMani,BIKemp1,BIBarn,
BIcankin2,BIMilB,BIKemp}.
We now know, as a result of the previous section, that {if} such a
decomposition into a component for the
field and a component for the motion of matter driven by a Lorentz force
exists then it can be uniquely formulated in terms of macroscopic
material property constants.
We simply need to determine what those property constants are.
\par
Gordon \cite{BIGord} began with the Lorentz force on a single electric
dipole with dipole moment ${\bf p}=\alpha {\bf e}$
\begin{equation}
{\bf f}_{dipole}=({\bf p}\cdot\nabla ){\bf e}
+\frac{d{\bf p}}{d(ct)}\times{\bf b}.
\label{EQvs.3.01}
\end{equation}
Here, ${\bf e}$ is the microscopic electric field,
${\bf b}$ is the microscopic magnetic field,
and ${\alpha}$ is the linear polarizability.
Assuming that the distance between dipoles in the material is much
smaller than the wavelength, then the theory can be considerably
simplified by applying an averaging scheme to a volume containing many
atoms per cubic radiation wavelength.
Gordon \cite{BIGord} describes an averaging process that he uses
to derive a macroscopic Lorentz force density
\begin{equation}
{\bf f}_L=N\langle{\bf f}_{dipole}\rangle
=\frac{N\alpha}{2}\left (
{\bf E}\times\frac{\partial{\bf B}}{\partial(ct)}
+\frac{\partial{\bf E}}{\partial(ct)}\times{\bf B}
\right ) \, ,
\label{EQvs.3.02}
\end{equation}
where $N$ is the number density,
${\bf E}=\langle{\bf e}\rangle$ is the macroscopic electric field,
and
${\bf B}=\langle{\bf b}\rangle$ is the macroscopic magnetic field.
Gordon \cite{BIGord} employs the low-density approximation
\begin{equation}
n=\sqrt{1+N\alpha}\approx 1+N\alpha/2
\label{EQvs.3.03}
\end{equation}
such that
\begin{equation}
{\bf f}_L=N\langle{\bf f}_{dipole}\rangle
\approx (n-1)\left (
{\bf E}\times\frac{\partial{\bf B}}{\partial(ct)}
+\frac{\partial{\bf E}}{\partial(ct)}\times{\bf B}
\right ) \, .
\label{EQvs.3.04}
\end{equation}
Other authors have derived variations of this basic relation.
Peierls \cite{BIPei} used the exact relation
\begin{equation}
n^2=1+N\alpha
\label{EQvs.3.05}
\end{equation}
to obtain
\begin{equation}
{\bf f}_L=N\langle{\bf f}_{dipole}\rangle
=\frac{n^2-1}{2}\left (
{\bf E}\times\frac{\partial{\bf B}}{\partial(ct)}
+\frac{\partial{\bf E}}{\partial(ct)}\times{\bf B}
\right ) \, .
\label{EQvs.3.06}
\end{equation}
Saldanha \cite{BISal} and Bradshaw, {\it et al.} \cite{BIBrad} find the
macroscopic Lorentz force
\begin{equation}
{\bf f}_L=
(n^2-1)
\left (
\nabla\left ( \frac{1}{2} E^2 \right )+
{\bf E}\times\frac{\partial{\bf B}}{\partial(ct)}
+\frac{\partial{\bf E}}{\partial(ct)}\times{\bf B}
\right ) .
\label{EQvs.3.07}
\end{equation}
If the Lorentz force, Eq.~(\ref{EQvs.3.07}), is separated into
components,
\begin{equation}
{\bf f}_{L1}=
(n^2-1) \nabla\left ( \frac{1}{2} E^2 \right ) 
\label{EQvs.3.08}
\end{equation}
\begin{equation}
{\bf f}_{L2}=
(n^2-1) \left ( 
{\bf E}\times\frac{\partial{\bf B}}{\partial(ct)}
+\frac{\partial{\bf E}}{\partial(ct)}\times{\bf B}
\right ) \, ,
\label{EQvs.3.09}
\end{equation}
then ${\bf f}_{L2}$ is in the form of the Abraham force and differs
from the Peierls result for the Lorentz force by a factor of 2. 
\par
We now have three different formulas for the macroscopic Lorentz
force, Eqs.~(\ref{EQvs.3.04}), (\ref{EQvs.3.06}), and (\ref{EQvs.3.09}),
that have been derived from the microscopic Lorentz force on a dipole.
Selecting one of these average Lorentz force densities to equate with
the macroscopic force density ${\bf f}_{mat}$, Eq.~(\ref{EQvs.2.06}),
will result in a value for the material parameter $\xi_e$ and
formulas for the field and material components of momentum in
a dielectric, Eqs.~(\ref{EQvs.2.07}) and (\ref{EQvs.2.08}).
Let us consider the consequences of using each of these formulas, in
turn.
Comparing the Gordon form of the macroscopic force density,
Eq.~(\ref{EQvs.3.04}), with Eq.~(\ref{EQvs.2.06}) we obtain
$\xi_e=n-1$.
Substituting this result into Eqs.~(\ref{EQvs.2.07}) and
(\ref{EQvs.2.08}) yields
\begin{equation}
{\bf G}_{field}=
\int_{\sigma} \frac{1}{c}({\bf E}\times{\bf B})dv
\label{EQvs.3.10}
\end{equation}
\begin{equation}
{\bf G}_{mat}=
\int_{\sigma} \frac{n-1}{c}({\bf E}\times{\bf B})dv,
\label{EQvs.3.11}.
\end{equation}
The Gordon form of the force, Eq.~(\ref{EQvs.3.04}), provides us with
the sensible result that the field momentum is the Abraham momentum.
But the Gordon form is an approximation.
Using Peierls's exact result, Eq.~(\ref{EQvs.3.06}),
in Eq.~(\ref{EQvs.2.06}), we obtain $\xi_e=(n^2-1)/2$.
As before, we substitute $\xi_e$ into
Eqs.~(\ref{EQvs.2.07}) and (\ref{EQvs.2.08}) and find that
\begin{equation}
{\bf G}_{field}=
\int_{\sigma} \frac{1+2n-n^2}{2c}({\bf E}\times{\bf B})dv
\label{EQvs.3.12}
\end{equation}
\begin{equation}
{\bf G}_{mat}=
\int_{\sigma} \frac{n^2-1}{2c}({\bf E}\times{\bf B})dv.
\label{EQvs.3.13}
\end{equation}
Then the Peierls result, Eq.~(\ref{EQvs.3.06}), leads to a field
momentum, Eq.~(\ref{EQvs.3.12}), that is negative for values of the
refractive index $n > 1+\sqrt{2}$.
Further, there is a factor of two discrepancy in the derivations of
Gordon \cite{BIGord} and Peierls \cite{BIPei} compared to the later
work \cite{BISal,BIBrad}.
Applying the same procedure to the final form of the Lorentz force,
Eq.~(\ref{EQvs.3.09}), results in
\begin{equation}
{\bf G}_{field}=
\int_{\sigma} \frac{1+n-n^2}{c}({\bf E}\times{\bf B})dv
\label{EQvs.3.14}
\end{equation}
\begin{equation}
{\bf G}_{mat}=
\int_{\sigma} \frac{n^2-1}{c}({\bf E}\times{\bf B})dv.
\label{EQvs.3.15}
\end{equation}
Then, the Lorentz force given by Eq.~(\ref{EQvs.3.09}) presents us again
with the problem of a negative field momentum for some values of $n$,
namely $n > (1+\sqrt{5})/2$.
Therefore, none of the versions of the Lorentz force that are reprised
here from the scientific literature are particularly appealing.
Clearly, this procedure to derive the macroscopic Lorentz force is
not sufficiently rigorous to obtain a result that is consistent with
the total momentum.
In the next section, we take a more detailed look at the microscopic
Lorentz force.
\par
\section{Lagrangian dynamics of the dielectric}
\par
The macroscopic force density, or Lorentz force, is
only significant where the amplitudes of fields are changing in time.
The effect of the macroscopic Lorentz force, Eq.~(\ref{EQvs.3.02}),
has been interpreted, in the context of a typical pulse
shape \cite{BIGord}.
There is an acceleration of the physical atoms/dipoles on the front
side of the pulse where the electromagnetic
field is increasing in strength \cite{BIGord}.
Then the atoms in the middle portion of the pulse travel at a generally
constant velocity until they are decelerated due to the decreasing
field strength on the trailing side of the pulse.
No momentum is left in the atoms after the pulse has passed.
The process has been described by saying that the material momentum
travels with the pulse \cite{BIGord}.
In this section, we derive the dynamics of the atoms/dipoles from the
Lagrangian.
\par
The standard Lagrangian for a charged particle with mass $m$ and 
charge $q$ located at a point ${\bf x}$ interacting with the
electromagnetic field is given
by \cite{BICohen}
$$
L=
\frac{1}{2}m{{\bf \dot x}^2}
+\frac{q}{c}\left ({\bf \dot x}\cdot{\bf A}({\bf x})
-\phi({\bf x})\right ) 
$$
\begin{equation}
+\int_{\sigma} \frac{1}{2}
\left (
\left (\frac{1}{c}\frac{\partial {\bf A}({\bf x})}{\partial t} +\nabla\phi({\bf x}) \right )^2 -(\nabla\times{\bf A}({\bf x}))^2
\right )
\, dv \, .
\label{EQps.4.01}
\end{equation}
Here, ${\bf A}$ is the vector potential, $\phi$ is the scalar potential and overdots denote partial differentiation with respect to time.
We take Eq.~(\ref{EQps.4.01}) as our given starting point.
\par
The derivation and properties of this equation, including the gauge invariance
properties, are discussed in Sec. II.B of Ref.~\cite{BICohen}.
We can construct a Lagrangian for a dipole interacting with the
field from a positive charge of mass $m_+$ located at ${\bf x}_+$
and a negative charge with mass $m_-$ located at ${\bf x}_-$ and
connected by a restoring constant $\kappa$
$$
L=
-\frac{1}{2} \kappa ( {\bf x}_+-{\bf x_-})^2
+\frac{1}{2}m_+{{\bf \dot x}_+^2}+\frac{1}{2}m_-{{\bf \dot x}_-^2}
$$
$$
+\frac{q}{c} \left ( {{\bf \dot x}_+}\cdot{\bf A}({{\bf x}_+})
-\phi({\bf x}_+)
- {{\bf \dot x}_-}\cdot{\bf A}({{\bf x}_-})
+\phi({\bf x}_-)
\right ) 
$$
\begin{equation}
+\int_{\sigma} \frac{1}{2}\left [\left (\frac{1}{c}\frac{\partial {\bf A}({\bf x})}{\partial t} +\nabla\phi({\bf x}) \right )^2 -(\nabla\times{\bf A}({\bf x}))\right ] \, dv \, .
\label{EQps.4.02}
\end{equation}
Transforming to a center-of-mass coordinate system, the Lagrangian is
$$
L=\left [ -\frac{1}{2}\kappa{\bf r}^2
+\frac{1}{2}M{\bf \dot R}^2
+\frac{1}{2}\mu{\bf \dot r}^2
+\frac{q}{c} {\bf \dot r}\cdot{\bf A}({\bf R})
\right ] 
$$
\begin{equation}
+\int_{\sigma} \frac{1}{2}\left [\left (\frac{1}{c}\frac{\partial {\bf A}({\bf x})}{\partial t} +\nabla\phi({\bf x}) \right )^2 -(\nabla\times{\bf A}({\bf x}))\right ] \, dv \, .
\label{EQps.4.03}
\end{equation}
where
$M=m_++m_-$ is the total mass,
$\mu=m_+m_-/M$ is the reduced mass,
${\bf r}={\bf x}_+-{\bf x}_-$ is the relative coordinate,
${\bf R}=(m_+{\bf x}_++m_-{\bf x}_-)/M$ is the center-of-mass
coordinate.
In writing Eq.~(\ref{EQps.4.03}), we have applied the usual dipole
approximation in which the electric field does not change much over the
distance between charges in the dipole \cite{BIWSQO}, such that
\begin{equation}
{\bf A}({{\bf x}_+}) \cong
{\bf A}({{\bf x}_-}) \cong
{\bf A}({\bf R})
\label{EQxx.4.0i}
\end{equation}
\begin{equation}
{\phi}({{\bf x}_+}) \cong
{\phi}({{\bf x}_-}) \cong
{\phi}({\bf R}) \,.
\label{EQxx.4.04}
\end{equation}
For each dipole in the material, the Lagrange equations of motion are
\begin{equation}
\frac{d}{dt}\left (\frac{\partial {L}}{\partial \dot {\bf R}_i} \right )
-\frac{\partial {L}}{\partial {\bf R}_i}
=0
\label{EQps.4.04}
\end{equation}
\begin{equation}
\frac{d}{dt}\left (\frac{\partial {L}}{\partial \dot {\bf r}_i} \right )
-\frac{\partial {L}}{\partial {\bf r}_i}
=0 \, .
\label{EQps.4.05}
\end{equation}
We will make use of the chain rule
\begin{equation}
\frac{d}{dt}
= \frac{\partial}{\partial t}
+\sum{\bf \dot R}_i\frac{\partial}{\partial {\bf R}_i}
+\sum{\bf \dot r}_i\frac{\partial}{\partial {\bf r}_i} \, .
\label{EQps.4.06}
\end{equation}
Applying Eq.~(\ref{EQps.4.04}) to the Lagrangian, Eq.~(\ref{EQps.4.03}),
we obtain a microscopic force equation 
\begin{equation}
{\bf f}_i= M{\bf \ddot R}_i
=\frac{q}{c}\frac{\partial}{\partial {\bf R}_i}
( {\bf \dot r} \cdot {\bf A}({\bf R} ))
\label{EQxx.4.06}
\end{equation}
for each component.
Combining the three vector components results in
\begin{equation}
{\bf f}= M{\bf \ddot R}
=\frac{q}{c}\nabla_R({\bf \dot r}\cdot{\bf A}({\bf R}))
\label{EQxx.4.86}
\end{equation}
where
\begin{equation}
\nabla_R=\left (\frac{\partial}{\partial R_x},
\frac{\partial}{\partial R_y},\frac{\partial}{\partial R_z} \right ) \,.
\label{EQps.4.07}
\end{equation}
Applying the vector identity
\begin{equation}
\nabla({\bf A}\cdot{\bf B})=
({\bf B}\cdot\nabla){\bf A}+
({\bf A}\cdot\nabla){\bf B}+
{\bf B}\times(\nabla\times{\bf A})+
{\bf A}\times(\nabla\times{\bf B})
\label{EQps.4.08}
\end{equation}
in the plane-wave limit, we obtain the microscopic force equation
\begin{equation}
{\bf f}_{dipole}=
M{\bf \ddot R}= \frac{q}{c}{\bf \dot r}\times(\nabla_R\times{\bf A}({\bf R}))
\label{EQps.4.09}
\end{equation}
that is the Lorentz force on a single dipole, that
is exerted at the center of mass.
In the plane-wave limit, the Lorentz force is in the direction of
propagation, or the opposite direction, such that $\nabla_R=\nabla$ without
loss of generality.
With definitions for the magnetic field
${\bf b}=\nabla\times {\bf A}$,
and the polarizability ${\bf p}=q{\bf r}$,
we have the Lorentz force on a dipole
\begin{equation}
{\bf f}_{dipole}=
M{\bf \ddot R}=
\frac{1}{c}{\bf \dot p}\times{\bf b}
\label{EQps.4.10}
\end{equation}
The Lorentz force on a dipole that is given by Eq.~(\ref{EQvs.3.01}) contains
an extra term $({\bf p}\cdot\nabla){\bf e}$ that does not appear in the 
Lorentz force, Eq.~(\ref{EQps.4.10}).
This term went out in the dipole approximation, which is fortunate because the
dipole force should not have a transverse component in the plane-wave limit.
\par
In the Gordon \cite{BIGord} model, the atoms are free particles that
are accelerated by the Lorentz force at the leading edge of the field
and travel at constant velocity until decelerated by the Lorentz force
at the trailing edge of the pulse.
Then, the momentum of the atoms travels with the field and contributes to
the total momentum.
However, that model assumes a low-density limit in which there is no
scattering of the field and assumes that the atoms move without being
restrained by association with a lattice and without colliding
with any other particles in the material, {\it i.e.} a vacuum.
Instead, there is substantial impediment to the motion of atoms in
a dielectric that, after all, is a material that can be modeled as
a continuous medium.
Then the canonical momentum
\begin{equation}
M{\bf \dot R}=\left (
\frac{\partial L}{\partial {\bf \dot R}_1},
\frac{\partial L}{\partial {\bf \dot R}_2},
\frac{\partial L}{\partial {\bf \dot R}_3}
\right ) 
\label{EQxx.4.12}
\end{equation}
makes no appreciable contribution to the total momentum.
\par
The other Lagrange equation of motion
\begin{equation}
\mu{\bf \ddot r} +\kappa{\bf r}= -\frac{q}{c}{\bf \dot A}
-\frac{q}{c} ({\bf \dot R}\cdot\nabla_r){\bf A}
\label{EQps.4.12}
\end{equation}
that is derived by applying Eq.~(\ref{EQps.4.05}) to the
Lagrangian, Eq.~(\ref{EQps.4.01}), and collecting the components is more
interesting because it is evocative of the other microscopic model of a
dielectric as stationary simple harmonic oscillators driven by the
electric field \cite{BIMVK}.
Setting ${\bf \dot R}=0$ in Eq.~(\ref{EQps.4.12}) we recover the
usual and familiar model of a dielectric as a collection of stationary
simple harmonic oscillators that are driven by the electric
field \cite{BIMVK}.
Defining a frequency $\omega_0=\sqrt{\kappa/\mu}$, the solution of 
\begin{equation}
{\bf \ddot p} +\omega_0^2{\bf p}
= -\frac{q^2}{\mu c}{\bf \dot A}
= \frac{q^2}{\mu }{\bf e}
\label{EQps.4.13}
\end{equation}
for a time harmonic field
${\bf e}=({\bf \tilde e}e^{-i(\omega_p t-kz)}+c.c.)/2$
is
\begin{equation}
{\bf \tilde p}=\frac{q^2/\mu}{\omega_0^2-\omega_p^2}{\bf \tilde e}
\label{EQps.4.14}
\end{equation}
Assuming the same relation holds for macroscopic fields, we have a 
macroscopic polarization field
\begin{equation}
{\bf \tilde P}=\frac{N q^2/\mu}{\omega_0^2-\omega_p^2}{\bf \tilde E} =\chi {\bf \tilde E}
\label{EQps.4.15}
\end{equation}
given a number density of dipoles $N$.
The fields, electric, magnetic, and polarization, travel together
as a pulse of electromagnetic radiation.
\par
It is well known that the polarization field contributes to the
electromagnetic energy 
\begin{equation}
U_{e} =
\int_{\sigma}\frac{1}{2}
\left (({\bf E}+{\bf P}){\bf E}+{\bf B}^2\right ) dv,
\label{EQps.4.16}
\end{equation}
as well as the electric and magnetic fields.
In terms of time-harmonic fields, the electromagnetic energy density is
\begin{equation}
\rho_{e} = n^2 \tilde E^2.
\label{EQps.4.17}
\end{equation}
Although a material may possess many forms of energy, we intend total
energy to include only those forms of energy that could impact the
dynamics or electrodynamics of the model system.
If the material is initially stationary in the laboratory frame of
reference, then it remains stationary with negligible kinetic energy
due to the gradient-index antireflection coating.
Then the electromagnetic energy is the total energy for a stationary
linear dielectric.
\par
Momentum is associated with the movement of energy.
The electromagnetic momentum will contain a contribution from the
movement of the polarization field, as well as contributions from the
movement of the electric and magnetic fields.
Writing the total momentum density, Eq.~(\ref{EQvs.1.03}),
in terms of time-harmonic fields, we have
\begin{equation}
{\bf g}_{total}
= \frac{n^2 \tilde E^2 }{c^2} c {\bf \hat e}_{\bf k}
= \frac{\rho_e }{c^2} c {\bf \hat e}_{\bf k} \, ,
\label{EQps.4.18}
\end{equation}
where ${\bf \hat e}_{\bf k}$ is a unit vector in the direction of
propagation.
Then, in a stationary dielectric medium, the equality of the total
momentum density and the electromagnetic energy density means that
the total momentum density, Eq.~(\ref{EQvs.1.03}), is equal to the
electromagnetic momentum density, Eq.~(\ref{EQvs.2.07}), that is,
${\bf G}_{field}={\bf G}_{total}$.
Then we have $\xi_e=0$ and substituting this result into
Eq.~(\ref{EQvs.2.08}), we find ${\bf G}_{mat}=0$.
There is no appreciable matter component of either energy or momentum
for a simple anti-reflection coated linear dielectric medium that is
at rest in the laboratory frame of reference.
\par
The situation changes somewhat in the absence of a gradient-index 
antireflection coating.
Momentum balance requires, at the cost of some amount of rigor in the
derivation, that we impute a momentum to the material that is twice
the momentum of the reflected field imparted by the surface force of
reflection.
But, that is the only significant material momentum, there is no
momentum of the Gordon type due to motion of the atoms inside the
dielectric.
\par
\section{Conclusions}
\par
The total momentum is uniquely conserved in a thermodynamically closed
system with complete equations of motion.
Once the total momentum is known, with certainty, from conservation
principles, then the momentum relations can be derived.
Using a microscopic model of the dielectric medium, we showed that
the propagating polarization field, not the movement of matter, is
a component of the momentum of a system consisting of a stationary
dielectric illuminated by a pulse of quasimonochromatic radiation
through a gradient-index antireflection coating.
\par
\vskip 2.718281828pt
\par

\end{document}